\documentclass[10pt,twocolumn,letterpaper]{article}

\usepackage[pagenumbers]{cvpr} 

\usepackage{graphicx}
\usepackage{amsmath}
\usepackage{amssymb}
\usepackage{booktabs}

\usepackage{epsfig}
\usepackage{cite}
\usepackage{verbatim}
\usepackage{color}
\usepackage{array}
\usepackage{epstopdf}
\usepackage{booktabs, multirow}       
\usepackage{stackengine}
\usepackage{colortbl}
\usepackage{multirow}
\usepackage{makecell}

\usepackage[pagebackref,breaklinks,colorlinks]{hyperref}

\usepackage[capitalize]{cleveref}
\crefname{section}{Sec.}{Secs.}
\Crefname{section}{Section}{Sections}
\Crefname{table}{Table}{Tables}
\crefname{table}{Tab.}{Tabs.}

\def\cvprPaperID{4515} 
\def\confName{CVPR}
\def\confYear{2022}

\begin{document}

\title{DPICT: Deep Progressive Image Compression Using Trit-Planes}

\author{Jae-Han Lee$^{1,2}$ \ \ \ \ \ Seungmin Jeon$^1$ \ \ \ \ \ Kwang Pyo Choi$^3$ \ \ \ \ \ Youngo Park$^3$ \ \ \ \ \ Chang-Su Kim$^1$\\
{\large $^1$Korea University \ \ \ \ \ $^2$Gauss Labs \ \ \ \ \ $^3$Samsung Electronics} \\
{\tt\small \{jaehanlee, seungminjeon\}@mcl.korea.ac.kr,} \\
{\tt\small \{kp5.choi, youngo.park\}@samsung.com, changsukim@korea.ac.kr}
}

\maketitle

\begin{abstract}
   We propose the deep progressive image compression using trit-planes (DPICT) algorithm, which is the first learning-based codec supporting fine granular scalability (FGS). First, we transform an image into a latent tensor using an analysis network. Then, we represent the latent tensor in ternary digits (trits) and encode it into a compressed bitstream trit-plane by trit-plane in the decreasing order of significance. Moreover, within each trit-plane, we sort the trits according to their rate-distortion priorities and transmit more important information first. Since the compression network is less optimized for the cases of using fewer trit-planes, we develop a postprocessing network for refining reconstructed images at low rates. Experimental results show that DPICT outperforms conventional progressive codecs significantly, while enabling FGS transmission. Codes are available at \href{https://github.com/jaehanlee-mcl/DPICT}{https://github.com/jaehanlee-mcl/DPICT}.
\end{abstract}

\section{Introduction}
\label{sec:introduction}

Image compression is a fundamental problem in image processing and analysis. Classical image codecs, such as JPEG \cite{y1992_CE_JPEG}, JPEG2000 \cite{y2001_SPM_JPEG2000}, WebP \cite{WebP}, and BPG \cite{bpg}, have been developed to achieve the goal of efficient data storage and transmission. They contain several modules to process hand-crafted features. For example, for transform coding, JPEG uses discrete cosine transform, and JPEG2000 adopts wavelet transform.

Recently, with the availability of substantial training data and computing resources, deep learning has been adopted for image compression, as well as other image and vision problems. Some learning-based codecs are based on convolutional neural networks (CNNs) \cite{y2017_ICLR_balle,y2018_ICLR_balle,y2018_NIPS_minnen,y2019_ICLR_lee}, while others on recurrent neural networks (RNNs) \cite{y2016_ICLR_toderici,y2016_NIPS_gregor,y2017_CVPR_toderici,y2018_CVPR_johnston}. In terms of the rate-distortion (RD) performance, recent learning-based codecs \cite{y2020_CVPR_cheng,y2021_CVPR_cui,y2021_CVPR_yang} are competitive with or even superior to the classical codecs.

\begin{figure}[t]
    \begin{center}
    \includegraphics[width=0.97\linewidth]{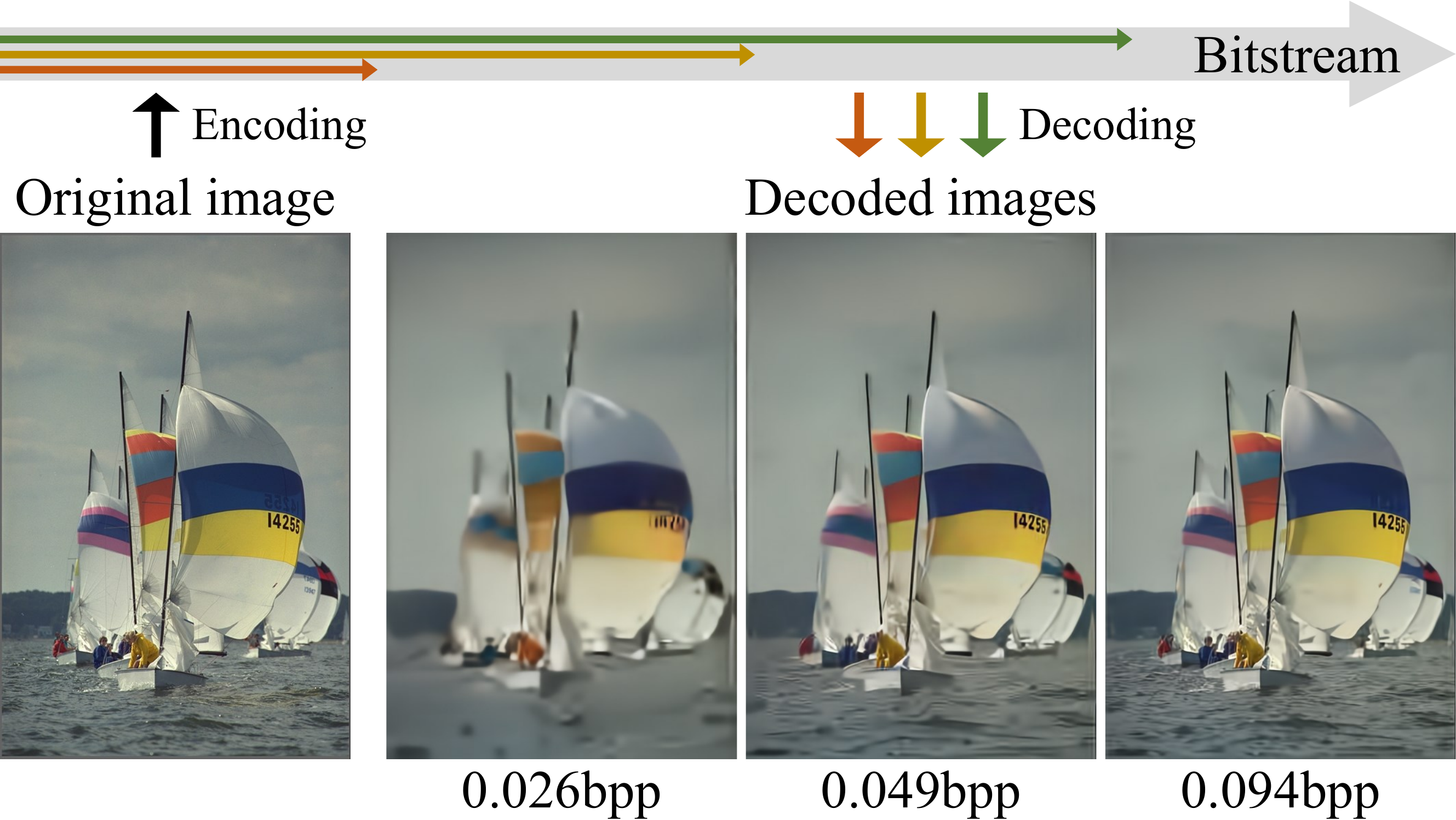}
    \end{center}
    \vspace*{-0.4cm}
    \caption
    {
        Illustration of progressive reconstruction of the proposed DPICT algorithm.
    }
    \label{fig:intro}
\end{figure}

Progressive compression, or scalable coding~\cite{y2005_IEEE_ohm}, is a crucial issue. A progressive codec encodes an image into a single bitstream that can decoded at various bit-rates, as illustrated in Figure~\ref{fig:intro}. There are various terminals from small wearable devices to big TVs, requiring different image qualities at different bit-rates. It is inefficient to encode multiple non-scalable bitstreams for these diverse devices. In contrast, a scalable bitstream can be efficiently truncated at multiple points to reconstruct the scene at different qualities. Moreover, when a network has a limited bandwidth, the receiver of a scalable bitstream can first check a preview image by receiving a small portion of the bitstream, and then reconstruct a higher quality image by decoding the remaining bits.

There are several learning-based progressive codecs~\cite{y2016_ICLR_toderici,y2016_NIPS_gregor,y2017_CVPR_toderici,y2018_CVPR_johnston,y2019_PCS_cai}, which, however, support coarse granular scalability only: a bitstream can be decoded at a limited number of rates. To the best of our knowledge, the proposed algorithm is the first learning-based codec with fine granular scalability (FGS) \cite{y1996_TCSVT_said,y2001_TCSVT_li}: a single bitstream can be truncated at any points to reconstruct the scene faithfully. Furthermore, despite this additional functionality, the proposed algorithm provides better RD performance than the conventional learning-based progressive codecs.

In this paper, we propose the deep progressive image codec using trit-planes (DPICT) with FGS. First, we transform an image into a latent tensor, each element of which is represented by ternary digits (trits). Then, we encode the latent tensor trit-plane by trit-plane in the decreasing order of significance. Moreover, even in the same trit-plane, we sort the trits according to the RD priorities to transmit more important information first. At the decoder, when fewer trit-planes are used, the reconstructed image is degraded by quantization errors and contain noisy artifacts. To reduce such artifacts, we also develop postprocessing networks. Experimental results demonstrate that DPICT outperforms the conventional progressive codecs significantly, while supporting FGS.

\section{Related Work}
\label{sec:related_work}

\vspace{0.1cm}
\noindent
\textbf{Learning-based compression:}
A typical learning-based image compression method \cite{y2017_ICLR_balle,y2017_ICLR_theis,y2017_NIPS_agustsson} constructs a neural network, by integrating a quantizer into an autoencoder \cite{y2008_ICML_vincent}, which is trained end-to-end to minimize a loss function, including rate and distortion terms. Given an image, the encoder (or analysis network) generates a latent representation, which is then quantized. The decoder (or synthesis network) dequantizes the representation and reconstructs the image in a lossy manner. The quantizer is, however, not differentiable. For the backpropagation in the training phase, it is approximated by the binarization process \cite{y2016_ICLR_toderici}, additive uniform noise \cite{y2017_ICLR_balle}, or stochastic rounding \cite{y2017_ICLR_theis}.

Rippel and Bourdev \cite{y2017_ICML_rippel}, Tschannen \etal \cite{y2018_NIPS_tschannen}, and Agustsson \etal \cite{y2019_ICCV_agustsson} adopted generative adversarial networks \cite{y2014_NIPS_goodfellow_GAN} for image compression. Nakanishi \etal \cite{y2018_ACCV_nakanishi} developed an image codec based on a multi-scale autoencoder. Ball{\'e} \etal \cite{y2018_ICLR_balle} proposed an additional autoencoder for a hyperprior. They assumed that latent elements have Gaussian distributions with zero mean, and used the hyperprior autoencoder to encode the standard deviations. Mentzer \etal \cite{y2018_CVPR_mentzer}, Minnen \etal \cite{y2018_NIPS_minnen}, Lee \etal \cite{y2019_ICLR_lee}, and Li \etal \cite{y2020_TIP_li} adopted context models. Also, more sophisticated hyperpriors have been studied. Cheng \etal \cite{y2020_CVPR_cheng} assumed Gaussian mixture models for latent elements, and Cui \etal \cite{y2021_CVPR_cui} used asymmetric Gaussian distributions.

\vspace{0.1cm}
\noindent
\textbf{Variable-rate compression:}
While traditional codecs, such as JPEG \cite{y1992_CE_JPEG}, support variable bit-rates, most learning-based codecs \cite{y2017_ICLR_balle,y2017_ICLR_theis,y2018_ICLR_balle,y2018_NIPS_minnen,y2019_ICLR_lee,y2020_CVPR_cheng,y2020_TIP_li} can generate bitstreams at fixed rates only. For multiple rates, they should train as many models, which incur inefficiency in testing, as well as in training.

Hence, several learning-based codecs have been developed to achieve variable-rate compression using a single network. Theis \etal \cite{y2017_ICLR_theis} proposed a variable-rate training scheme for a single autoencoder using a scale parameter for quantization.
Choi \etal \cite{y2019_ICCV_choi} also employed the scaled quantization, while training rate-specific parameters for a few selected rates.
Similarly, Yang \etal \cite{y2020_SPL_yang} proposed the modulated autoencoders containing separate modules for selected rates. Cai \etal \cite{y2018_TCSVT_cai} generated multi-scale representations and performed content-adaptive rate allocation.
Cui \etal \cite{y2021_CVPR_cui} proposed gain units to guide the network to allocate more bits to specific channels and also to control rates. Yang \etal \cite{y2021_CVPR_yang} used the slimmable neural networks \cite{y2018_ICLR_yu,y2019_ICCV_yu} to perform low-rate compression using only a fraction of the parameters and the highest-rate compression using all parameters. These codecs \cite{y2017_ICLR_theis, y2019_ICCV_choi, y2020_SPL_yang, y2018_TCSVT_cai, y2021_CVPR_cui, y2021_CVPR_yang} can adapt to different rates with a single trained model, but their bitstreams are not scalable, \ie a lower-rate bitstream is not embedded in a higher-rate one.

\vspace{0.1cm}
\noindent
\textbf{Progressive compression:} Progressive codecs to encode scalable bitstreams have also been studied. For example, JPEG \cite{y1992_CE_JPEG} and JPEG2000 \cite{y2001_SPM_JPEG2000} are traditional progressive codecs. There are several learning-based progressive codecs, most of which are based on RNNs \cite{y2016_ICLR_toderici,y2016_NIPS_gregor,y2017_CVPR_toderici,y2018_CVPR_johnston}. Toderici \etal \cite{y2016_ICLR_toderici} proposed the first RNN-based progressive codec. Their network, utilizing the long short-term memory (LSTM) \cite{y1997_NC_hochreiter_LSTM}, transmits bits progressively: At stage $t+1$, the encoder transmits the residual error at stage $t$, and the decoder reconstructs it and adds it to the reconstructed image at stage $t$. If this is repeated $T$ times, the single bitstream can support $T$ different rates progressively. Gregor \etal \cite{y2016_NIPS_gregor} also introduced a recurrent image codec, which improves conceptual quality using a generative model. However, these codecs \cite{y2016_ICLR_toderici,y2016_NIPS_gregor} are for low-resolution patches. Torderici \etal \cite{y2017_CVPR_toderici} developed a codec for higher-resolution images by expanding the previous work in \cite{y2016_ICLR_toderici}. Johnston \etal \cite{y2018_CVPR_johnston} proposed an effective initializer for hidden states of their RNN and a spatially adaptive rate controller. However, all these RNN-based algorithms support only coarse granular scalability: a bitstream can be reconstructed at $T$ different rates only, where $T$ is the number of recurrent stages. Moreover, their compression performances are inferior even to those of the traditional codecs \cite{bpg}. On the other hand, Cai \etal \cite{y2019_PCS_cai} proposed a network of a single encoder and two decoders. The encoder decomposes an image into two representations. Then, the preview decoder uses only one representation, and the full-quality decoder uses both representations. Hence, their network supports two rates only.

The proposed DPICT algorithm supports FGS in contrast to these learning-based progressive codecs\cite{y2016_ICLR_toderici,y2016_NIPS_gregor,y2017_CVPR_toderici,y2018_CVPR_johnston,y2019_PCS_cai}. Furthermore, DPICT provides significantly better RD performance than the conventional progressive codecs.

\begin{figure}[t]
    \begin{center}
    \includegraphics[width=0.97\linewidth]{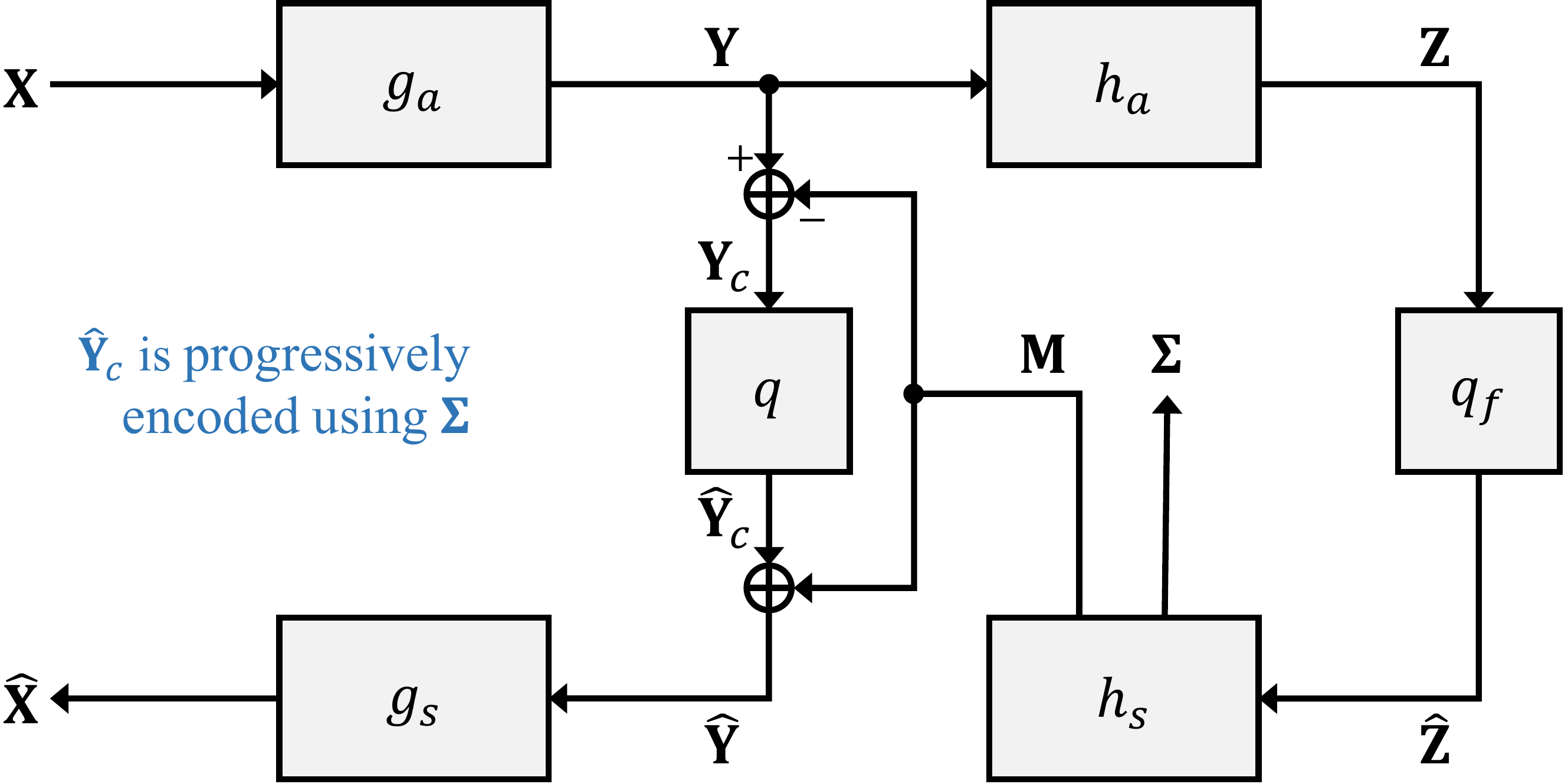}
    \end{center}
    \vspace*{-0.4cm}
    \caption
    {
        Image compression framework.
    }
    \label{fig:model_schematic_1}
\end{figure}

\section{Proposed Algorithm}
\label{sec:proposed_algorithm}

\subsection{Compression network}
We adopt the compression framework in Figure~\ref{fig:model_schematic_1}, which consists of an encoder $g_a$, a decoder $g_s$, a hyper encoder $h_a$, and a hyper decoder $h_s$, as done in \cite{y2018_ICLR_balle,y2020_CVPR_cheng,y2019_ICCV_choi,y2021_CVPR_cui,y2019_ICLR_lee,y2018_NIPS_minnen, y2021_CVPR_yang}. An image $\mathbf{X}$ is transformed to a latent representation $\mathbf{Y}$ and a hyper latent representation $\mathbf{Z}$ sequentially by $g_a$ and $h_a$. Using the factorized prior model \cite{y2018_ICLR_balle}, denoted by $q_f(\cdot)$ in Figure~\ref{fig:model_schematic_1}, $\mathbf{Z}$ is digitized to $\hat{\mathbf{Z}}$. From $\hat{\mathbf{Z}}$, $h_s$ yields $\mathbf{M}$ and $\mathbf{\Sigma}$, which contain the means and standard deviations of the elements in $\mathbf{Y}$, respectively. These elements are assumed to be independent Gaussian random variables. Then, the mean-removed (or centered) $\mathbf{Y}_{c} = \mathbf{Y} - \mathbf{M} $ is quantized to
\begin{equation}
\hat{\mathbf{Y}}_{c} = q(\mathbf{Y}_{c})
\label{eq:coded_data}
\end{equation}
where rounding is used for the quantizer $q(\cdot)$.
Finally, the decoder $g_s$ adds $\mathbf{M}$ back to $\hat{\mathbf{Y}}_{c}$ to yield
\begin{equation}
\hat{\mathbf{Y}} = \hat{\mathbf{Y}}_{c} + \mathbf{M}
\end{equation}
and uses it to reconstruct $\hat{\mathbf{X}}$.

In addition to $\hat{\mathbf{Z}}$, $\hat{\mathbf{Y}}_{c}$ in \eqref{eq:coded_data} is encoded into a bitstream. Let $\hat{y}_{c}$, $y_{c}$, and $\sigma$ be corresponding elements in $\hat{\mathbf{Y}}_{c}$, $\mathbf{Y}_{c}$, and $\mathbf{\Sigma}$. Then, the number of bits for encoding $\hat{y}_{c}$ is given by
\begin{equation}
\textstyle
N(\hat{y}_{c}) = - \log_2 P\left(\hat{y}_{c}-\frac{1}{2} \leq y_{c} < \hat{y}_{c}+\frac{1}{2} \right)
\label{eq:estimated_bit_rate}
\end{equation}
where $y_{c} \sim \mathcal{N}(0, \sigma^2)$. Unlike conventional algorithms, we compress $\hat{\mathbf{Y}}_{c}$ progressively using trit-planes in Section~\ref{ssec:trit_plane_coding}.

In the training phase, since the quantizer is not differentiable, it is approximated by an additive noise function $u(\cdot)$,
\begin{equation}
\textstyle
\tilde{\mathbf{Y}} = u(\mathbf{Y}) = \mathbf{Y} + \mathcal{U}\left(-\frac{1}{2}, \frac{1}{2}\right)
\label{eq:model1_train}
\end{equation}
where $\mathcal{U}(-\frac{1}{2}, \frac{1}{2})$ is a uniform noise tensor in range $(-\frac{1}{2}, \frac{1}{2})$. The noise function $u(\cdot)$ is assumed to generate white noise and not considered during the back-propagation of gradients \cite{y2017_ICLR_balle}. Similarly, $\tilde{\mathbf{Z}}$ is obtained from $\mathbf{Z}$~\cite{y2018_NIPS_minnen}, and then $\tilde{\mathbf{M}}$ and $\tilde{\mathbf{\Sigma}}$ are estimated. Finally, $\tilde{\mathbf{X}}$ is reconstructed from $\tilde{\mathbf{Y}}$.

The loss function $\ell$ consists of a distortion term $\ell_{D}$ and a rate term $\ell_{R}$,
\begin{equation}
\ell =  \ell_{D} ( \mathbf{X}, \tilde{\mathbf{X}} ) + \lambda \cdot \ell_{R}( \tilde{\mathbf{Y}}, \tilde{\mathbf{Z}}; \tilde{\mathbf{M}}, \tilde{\mathbf{\Sigma}} )
\label{eq:loss_function}
\end{equation}
where $\ell_{D}$ is defined as the mean squared error between $\mathbf{X}$ and $\tilde{\mathbf{X}}$, and $\ell_{R}$ is the rate for encoding the elements in $\tilde{\mathbf{Y}}$ and $\tilde{\mathbf{Z}}$. Notice that the rate for $\tilde{\mathbf{Y}}$ is given by the sum of the numbers of bits $N(\tilde{y})$, similar to \eqref{eq:estimated_bit_rate}, but based on $\tilde{\mathbf{M}}$ and $\tilde{\mathbf{\Sigma}}$. In \eqref{eq:loss_function}, $\lambda$ is a Lagrangian multiplier to control the RD trade-off.

\begin{figure}[t]
    \begin{center}
    \includegraphics[width=0.97\linewidth]{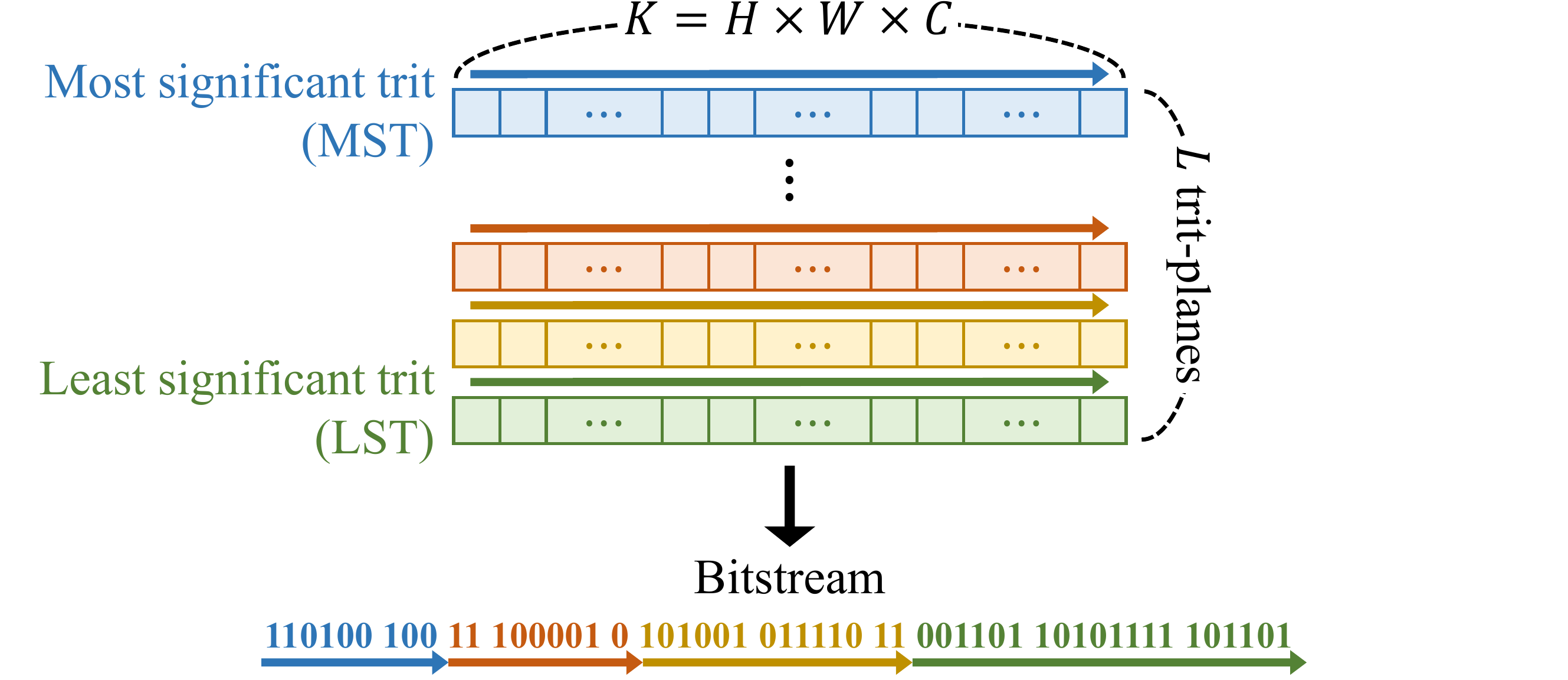}
    \end{center}
    \vspace*{-0.4cm}
    \caption
    {
        Illustration of the trit-plane coding of $K$ elements.
    }
    \label{fig:progressive_codec}
\end{figure}

\subsection{Trit-plane coding}
\label{ssec:trit_plane_coding}

We represent $\mathbf{\hat{Y}}_{c}$ in \eqref{eq:coded_data} in a ternary number system and encode it trit-plane by trit-plane to achieve FGS. There are $K=H\times W \times C$ elements in $\mathbf{\hat{Y}}_{c}$, where $H$, $W$, and $C$ are the height, the width, and the number of channels, respectively. Conventional algorithms transmit these $K$ elements in a raster scan order, so they cannot decompress a partial bitstream meaningfully; the reconstructed image is severely degraded if some of the elements are missing. In contrast, we express each element using $L$ trits and compress the $L$ trit-planes in the decreasing order of significance from the most significant trit (MST) to the least significant trit (LST), as shown in Figure~\ref{fig:progressive_codec}. In this way, front parts of the bitstream contain more important information, enabling the decoder to perform progressive reconstruction faithfully.

\begin{figure}[t]
    \begin{center}
    \includegraphics[width=0.97\linewidth]{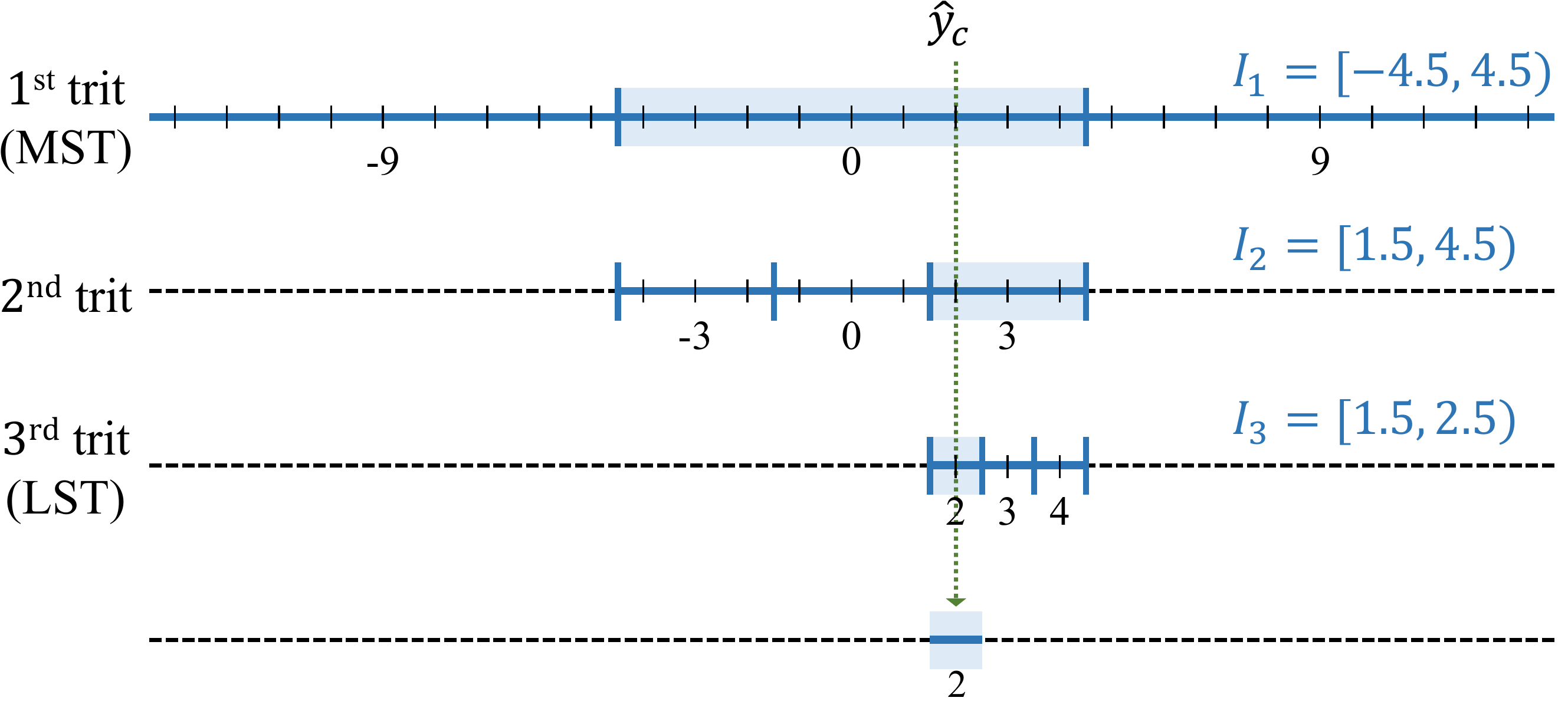}
    \end{center}
    \vspace*{-0.4cm}
    \caption
    {
        Trit-plane slicing of element $\hat{y}_c$.
    }
    \label{fig:trit_plane_coding}
\end{figure}

Figure \ref{fig:trit_plane_coding} illustrates how to progressively compress an element $\hat{y}_{c}=q(y_{c})$ in $\mathbf{\hat{Y}}_{c}$, where $y_{c} \sim \mathcal{N}(0, \sigma^2)$. In this example, $L=3$, so it is assumed that $y_c$ is rounded to one of the integers between $-13$ and $13=\frac{3^L-1}{2}$. First, the number line is partitioned into three intervals, and the MST indicates which interval contains $y_c$. Specifically, $0_{(3)}$, $1_{(3)}$, and $2_{(3)}$ correspond to the left, middle, and right intervals, respectively. In Figure~\ref{fig:trit_plane_coding}, $y_c$ belongs to the middle interval ${\cal I}_1 = [-4.5, 4.5)$, so the trit $1_{(3)}$ is encoded into the bitstream. Second, ${\cal I}_1 = [-4.5, 4.5)$ is partitioned into three sub-intervals, and the next trit $2_{(3)}$ informs that $y_c \in {\cal I}_2 = [1.5, 4.5)$. Finally, ${\cal I}_2$ is partitioned again, and the third trit (LST) $0_{(3)}$ means that $y_c \in {\cal I}_3 = [1.5, 2.5)$. Hence, with all three trits, the decoder knows that $\hat{y}_{c}=2$.

In general, let ${\cal I}_n = [l_n, r_n)$ denote the interval containing $y_c$ when the first $n$ trits are encoded. Also, let $t_n \in \{0_{(3)}, 1_{(3)}, 2_{(3)} \}$ denote the $n$th trit. ${\cal I}_{n}$ is partitioned into three sub-intervals of the same length (except for the leftmost and rightmost intervals):
\begin{align}
\textstyle
{\cal I}_{n}^0 &= [l_n^0, r_n^0) = [l_n, \textstyle \frac{2l_n + r_n}{3}),  \\
{\cal I}_{n}^1 &= [l_n^1, r_n^1) = [\textstyle \frac{2l_n + r_n}{3}, \textstyle \frac{l_n + 2r_n}{3}), \\
{\cal I}_{n}^2 &= [l_n^2, r_n^2) = [\textstyle \frac{l_n + 2r_n}{3}, r_n).
\end{align}
Then, the next $(n+1)$th trit $t_{n+1}$ reveals which subinterval contains $y_c$. It becomes ${\cal I}_{n+1}$.

The conditional probability of $t_{n}$ given $\{t_{n-1}, \cdots, t_1\}$ is given by
\begin{align}
\label{eq:cond_prob}
P(t_{n}|t_{n-1}, \cdots, t_1) &= P(y_c \in {\cal I}_{n} | y_c \in {\cal I}_{n-1}) \\
& =  \textstyle \frac{\Phi(r_n/\sigma) - \Phi(l_n/\sigma)}{\Phi(r_{n-1}/\sigma) - \Phi(l_{n-1}/\sigma)}
\end{align}
where $\Phi(\cdot)$ is the CDF of the standard normal distribution. The number of bits for encoding $t_n$ is then given by
\begin{equation}
N(t_n) = - \log_2 P(t_n|t_{n-1}, \cdots, t_1).
\label{eq:estimated_bit_rate_conditional}
\end{equation}
Also, by the chain rule,
\begin{equation}
\textstyle
P(\hat{y}_c) = P(t_L, \cdots, t_1) = \prod_{n=1}^{L} P(t_n|t_{n-1}, \cdots, t_1).
\label{eq:chain_rule}
\end{equation}
From \eqref{eq:estimated_bit_rate}, \eqref{eq:estimated_bit_rate_conditional}, and \eqref{eq:chain_rule}, we have
\begin{equation}
\textstyle
N(\hat{y}_c) = \sum_{n=1}^{L} N(t_n).
\label{eq:bit_equivalence}
\end{equation}
In other words, the trit-plane coding of $\hat{y}_c$ requires the same number of bits as the straightforward coding does. However, it allows progressive reconstruction of $y_c$. Suppose that the first $n$ trits are received. Then, the decoder reconstructs $y_c$ to
\begin{equation}
\hat{y}_c^n = E[y_c|y_c \in {\cal I}_n],
\label{eq:reconstruction_level}
\end{equation}
which is the minimum mean squared error (MMSE) estimate~\cite{y1991_gersho}. In other words, $\hat{y}_c^n$ in \eqref{eq:reconstruction_level} yields the minimum mean squared distortion
\begin{equation}
\label{eq:distortion}
D_n = E[(y_c - \hat{y}_c^n)^2|y_c \in {\cal I}_n].
\end{equation}

\vspace{0.1cm}
\noindent
\textbf{Bit-plane coding:}
Bit-planes may be used instead of trit-planes. However, note that the most frequent $\hat{y}_c$ is 0 because $y_{c} \sim \mathcal{N}(0, \sigma^2)$. When $\hat{y}_c$ is 0, $\hat{y}_c^n$ in \eqref{eq:reconstruction_level} is 0 regardless of $n$, which enables faithful reconstruction even at a low bit-rate. This is impossible in the bit-plane coding, \ie $\hat{y}_c^n \neq 0$ for $n \leq L-1$ when $\hat{y}_c = \hat{y}_c^L = 0$. Hence, we use trit-planes. It is shown in Section~\ref{sec:experiments} that the trit-plane coding outperforms the bit-plane coding.

\vspace{0.1cm}
\noindent
\textbf{Coding of $\mathbf{\hat{Z}}$:}
We do not compress $\mathbf{\hat{Z}}$ progressively. Since the entropy coding of $\hat{y}$ depends on $\mu$ and $\sigma$ estimated from $\mathbf{\hat{Z}}$, it is impossible to decompress $\hat{y}$ from a partially reconstructed $\mathbf{\hat{Z}}$. Also, $\mathbf{\hat{Z}}$ occupies only about 1\% of the overall bitrate. Hence, we transmit all bits for $\mathbf{\hat{Z}}$ before the progressive transmission of $\hat{\mathbf{Y}}_c$.

\subsection{RD-prioritized transmission}
\label{ssec:rd_prioritized_transmission}
To transmit more important information first, we compress the $L$ trit-planes from MST to LST. Moreover, within each trit-plane, we transmit the $K$ trits after sorting them according to their RD priorities. The transmission of a trit increases the rate ($\Delta R > 0$), but it decreases the distortion ($\Delta D < 0$). The goal is to minimize $\Delta R$ and maximize $-\Delta D$ simultaneously. To achieve this goal, sophisticated RD methods \cite{y1998_SPM_ortega} can be applied. However, for simplicity, we use the ratio $- \frac{\Delta D}{\Delta R}$ for RD-prioritized transmission.

\begin{figure}[t]
    \begin{center}
    \includegraphics[width=0.97\linewidth]{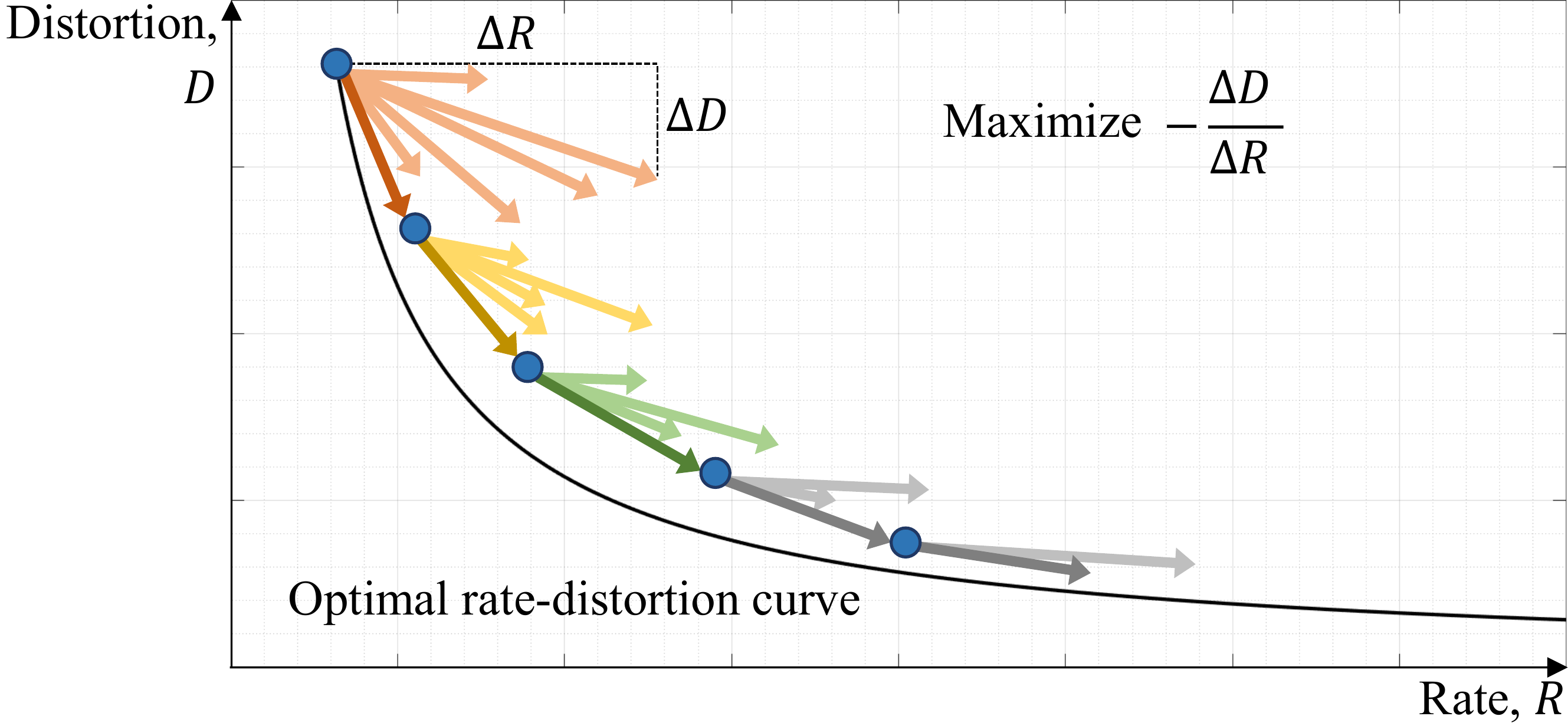}
    \end{center}
    \vspace*{-0.4cm}
    \caption
    {
       Greedy RD optimization using ratios $- \frac{\Delta D}{\Delta R}$.
    }
    \label{fig:compression_priority}
    \vspace*{0.1cm}
\end{figure}

\begin{figure}[t]
    \begin{center}
    \includegraphics[width=\linewidth]{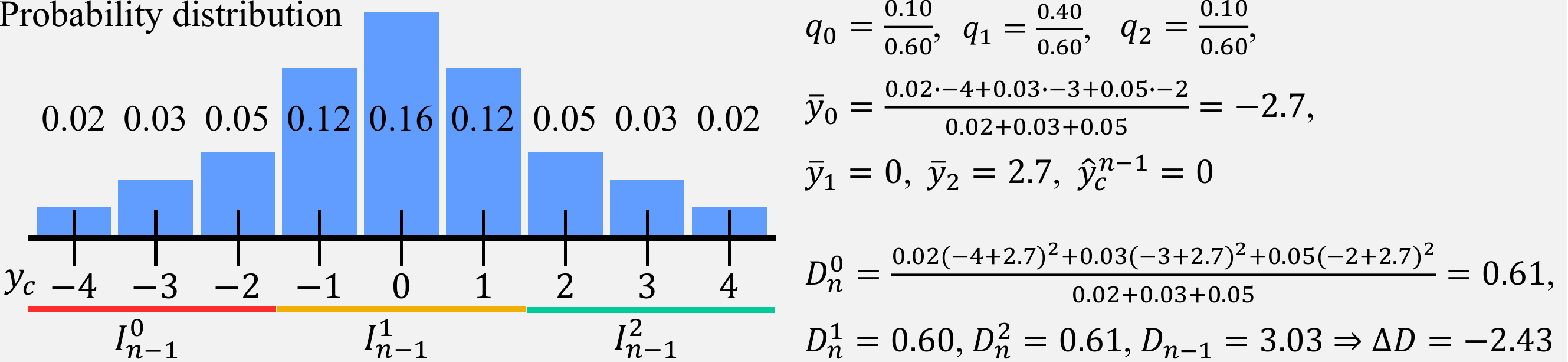}
    \end{center}
    \vspace*{-0.5cm}
    \caption
    {
        Toy example for computing $\bar{y}_k$, $D_n^k$, and $\Delta D$.
    }
    \label{fig:r2_example}
\end{figure}

In Figure~\ref{fig:compression_priority}, from the upper-left dot, there are several arrows corresponding to possible coding elements. We attempt to follow the optimal RD curve by selecting the element with the maximum ratio $- \frac{\Delta D}{\Delta R}$. After transmitting it, we repeat the process with the remaining elements. It is computationally prohibitive to measure $\Delta D$ exactly. Thus, we assume that the image distortion is proportional to the error in a latent element. The greedy selection and the distortion assumption cannot guarantee optimal transmission, but they yield good RD performance in practice, as will be shown in Section~\ref{sec:experiments}.

Suppose that the $n$th trit-plane is to be compressed. Thus, for an element $\hat{y}_c$, its first $n-1$ trits are already transmitted and its $n$th trit $t_n$ is to be compressed. Both the encoder and the decoder can compute the probabilities $q_k = P(t_n=k | t_{n-1}, \cdots, t_1)$, $k=0, 1, 2$ via \eqref{eq:cond_prob}. Then, the expected number of bits for encoding $t_n$ is given by the entropy $H(\{q_0, q_1, q_2\})$,
\begin{equation}
\textstyle
\Delta R = H(\{q_0, q_1, q_2\}) = - \sum_{k=0}^{2} q_k \log_2 q_k.
\label{eq:DeltaR}
\end{equation}
We use the ANS coder \cite{y2013_arXiv_duda_ANS} for the entropy coding. Also, for the three cases of $t_n = 0, 1, 2$, we compute the distortions via \eqref{eq:distortion}, respectively, which are denoted by $D_n^0, D_n^1, D_n^2$. Then, the expected distortion change is given by
\begin{equation}
\textstyle
\Delta D = E[D_{n}]-D_{n-1}= \sum_{k=0}^{2} q_k D_n^k - D_{n-1}
\label{eq:DeltaD}
\end{equation}
where $D_n^k$ is
\begin{equation}
\textstyle
D_n^k = E[(y_c - \bar{y}_k)^2|y_c \in {\cal I}_{n-1}^k]
\label{eq:subintervalD}
\end{equation}
with $\bar{y}_k = E[y_c|y_c \in {\cal I}_{n-1}^k]$. Figure~\ref{fig:r2_example} illustrates how to compute $\bar{y}_k$, $D_n^k$, and $\Delta D$ with a toy example.

Finally, we compute the RD priority of the $n$th trit, which is defined as
\begin{equation}
\label{eq:priority}
- \frac{\Delta D}{\Delta R} = \frac{\sum_{k=0}^{2} q_k D_n^k - D_{n-1}}{\sum_{k=0}^{2} q_k \log_2 q_k}.
\end{equation}
Then, we transmit the $K$ trits in each trit-plane after sorting them in the decreasing order of their RD priorities. In this way, more important information is transmitted first, even in the same trit-plane.

\begin{figure}[t]
    \begin{center}
    \includegraphics[width=0.97\linewidth]{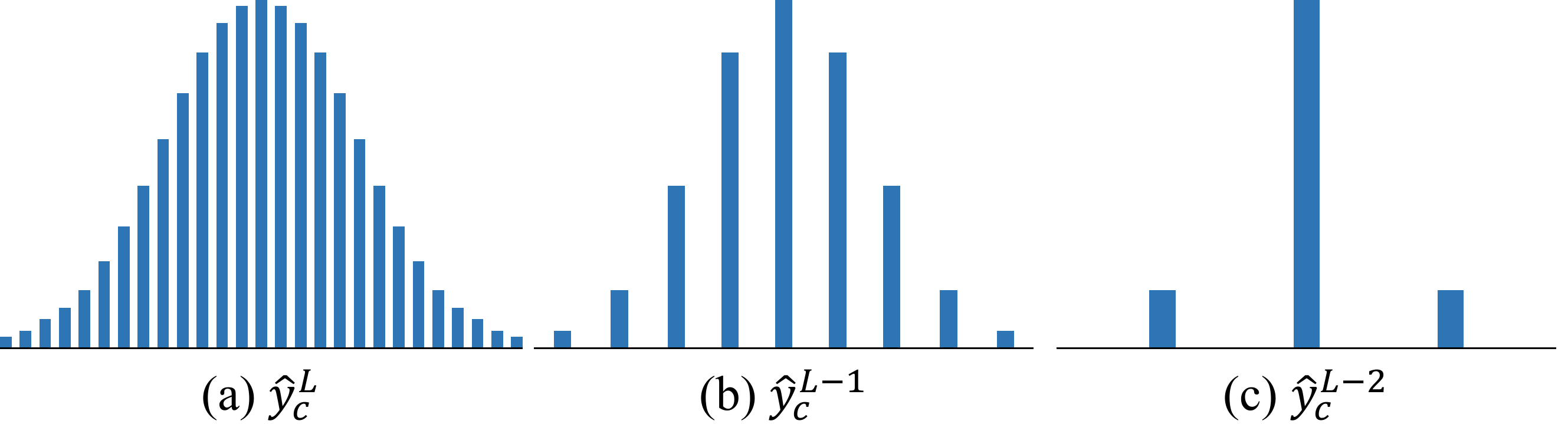}
    \end{center}
    \vspace*{-0.5cm}
    \caption
    {
        Histograms of (a) $\hat{y}_c^{L}$, (b) $\hat{y}_c^{L-1}$, and (c) $\hat{y}_c^{L-2}$.
    }
    \label{fig:quantized_gaussian}
\end{figure}

\begin{figure}[t]
    \begin{center}
    \includegraphics[width=0.97\linewidth]{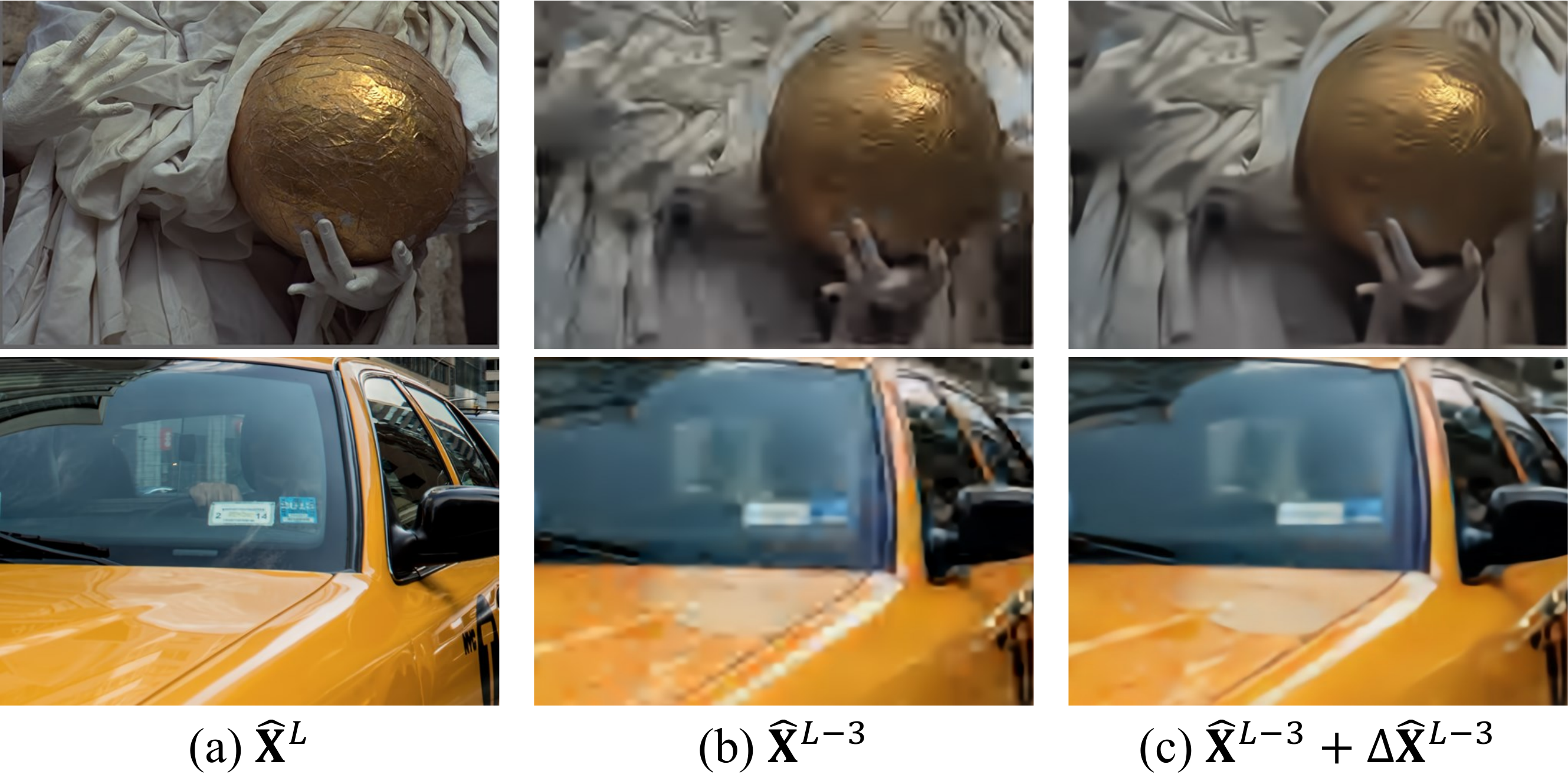}
    \end{center}
    \vspace*{-0.5cm}
    \caption
    {
        Reconstructed images (a) using all trit-planes and (b) using three fewer trit-planes. (c) Refined result of (b).
    }
    \label{fig:decompressed_low_trits}
\end{figure}

\subsection{Postprocessing network}
To train the proposed algorithm, in \eqref{eq:model1_train}, we use a uniform noise tensor in range $(-\frac{1}{2}, \frac{1}{2})$ to consider the case of using all $L$ trit-planes. Thus, the network is less optimized for the cases of using fewer trit-planes. Figure~\ref{fig:quantized_gaussian} shows histograms of $\hat{y}_c^{L}$, $\hat{y}_c^{L-1}$, and $\hat{y}_c^{L-2}$. With fewer trit-planes, the gaps between reconstruction levels get wider, resulting in bigger quantization errors, which generate more artifacts in reconstructed images. Let $\hat{\mathbf{X}}^n$ denote a reconstructed image from a partial representation $\hat{\mathbf{Y}}_c^n$ consisting of $\hat{y}_c^n$'s. As shown in Figure \ref{fig:decompressed_low_trits}(b), $\hat{\mathbf{X}}^{L-3}$ contains noisy artifacts.

\begin{figure}[t]
    \begin{center}
    \includegraphics[width=0.97\linewidth]{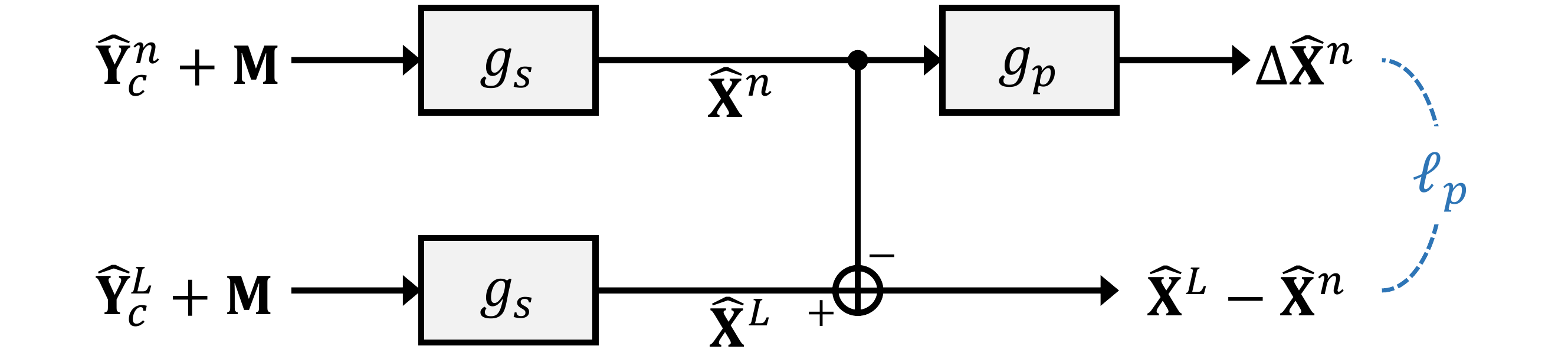}
    \end{center}
    \vspace*{-0.4cm}
    \caption
    {
        Training schema of a postprocessing network $g_p$.
    }
    \label{fig:addon_decoder_a}
\end{figure}

We develop a postprocessing network $g_p$ to reduce such artifacts. In other words, the goal of $g_p$ is to convert a reconstructed image $\hat{\mathbf{X}}^n$ to a more faithful one $\hat{\mathbf{X}}^n + \Delta \hat{\mathbf{X}}^n$ with less artifacts. Figure~\ref{fig:addon_decoder_a} shows the training schema of $g_p$. In the upper pass, we first obtain $\hat{\mathbf{Y}}_c^n$ by \eqref{eq:reconstruction_level} and then reconstruct $\hat{\mathbf{X}}^n=g_s(\hat{\mathbf{Y}}_c^n + \mathbf{M})$. Similarly, in the lower pass, we obtain the reconstructed image $\hat{\mathbf{X}}^L$ from $\hat{\mathbf{Y}}_c^L + \mathbf{M}$ using all trit-planes. We regard $\hat{\mathbf{X}}^L$ as a clean image without artifacts. Then, we train the postprocessing network to take $\hat{\mathbf{X}}^n$ and yield the residual $\Delta \hat{\mathbf{X}}^n$, by employing the postprocessing loss
\begin{equation}
\textstyle
\ell_p = \| \Delta \hat{\mathbf{X}}^n - (\hat{\mathbf{X}}^L - \hat{\mathbf{X}}^n) \|_2
\end{equation}
In the testing phase, using $g_p$, we refine a reconstructed image $\hat{\mathbf{X}}^n$ into $\hat{\mathbf{X}}^n + g_p(\hat{\mathbf{X}}^n)$. Figure \ref{fig:decompressed_low_trits}(c) shows how $g_p$ improves the reconstructed images $\hat{\mathbf{X}}^{L-3}$. We see that, compared to $\hat{\mathbf{X}}^{L-3}$, the refined images $\hat{\mathbf{X}}^{L-3} + \Delta \hat{\mathbf{X}}^{L-3}$ contain less artifacts and render the scenes more faithfully.

\begin{figure*}[h]
    \begin{center}
    \includegraphics[width=\linewidth]{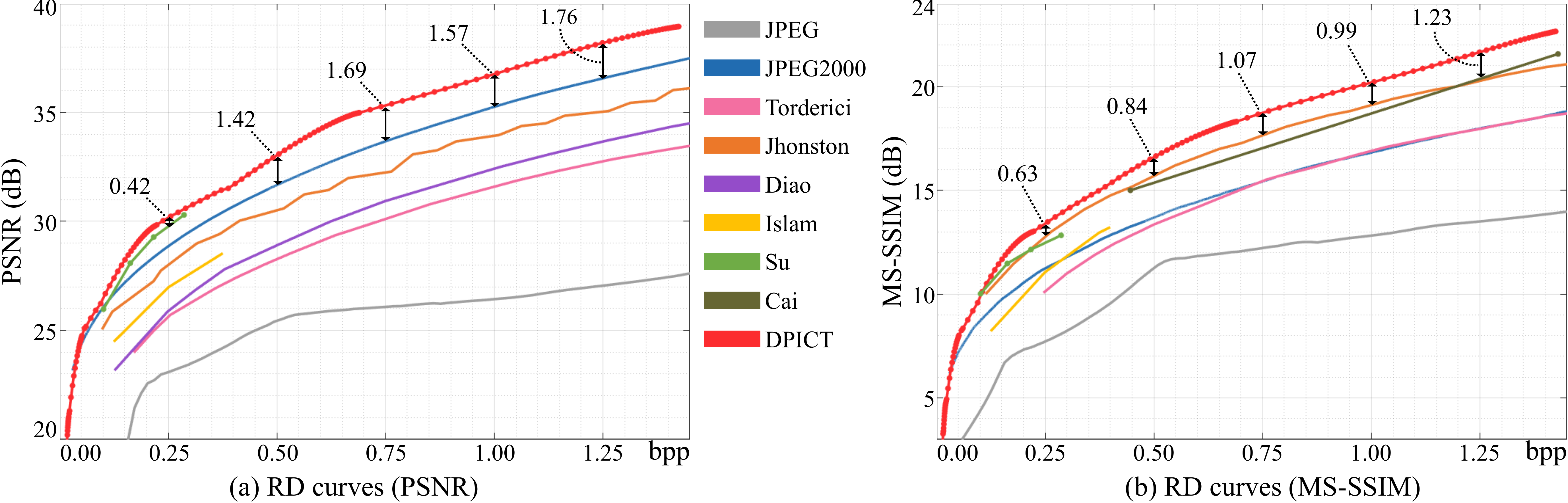}
    \end{center}
    \vspace*{-0.4cm}
    \caption
    {
        RD performance comparison of the proposed DPICT with conventional \textit{progressive} codecs on the Kodak dataset: JPEG~\cite{y1992_CE_JPEG}, JPEG2000~\cite{y2001_SPM_JPEG2000}, Torderici~\etal~\cite{y2017_CVPR_toderici}, Jhonston~\etal~\cite{y2018_CVPR_johnston}, Diao~\etal~\cite{y2020_DOC_diao}, Islam~\etal~\cite{y2021_CVPRW_islam}, Su~\etal~\cite{y2020_ICIP_su}, and Cai~\etal~\cite{y2019_PCS_cai}. At selected rates, the performance gaps between DPICT and the second best codecs are specified.
    }
    \label{fig:RD_curve_main}
\end{figure*}

\begin{figure}[t]
    \begin{center}
    \includegraphics[width=\linewidth]{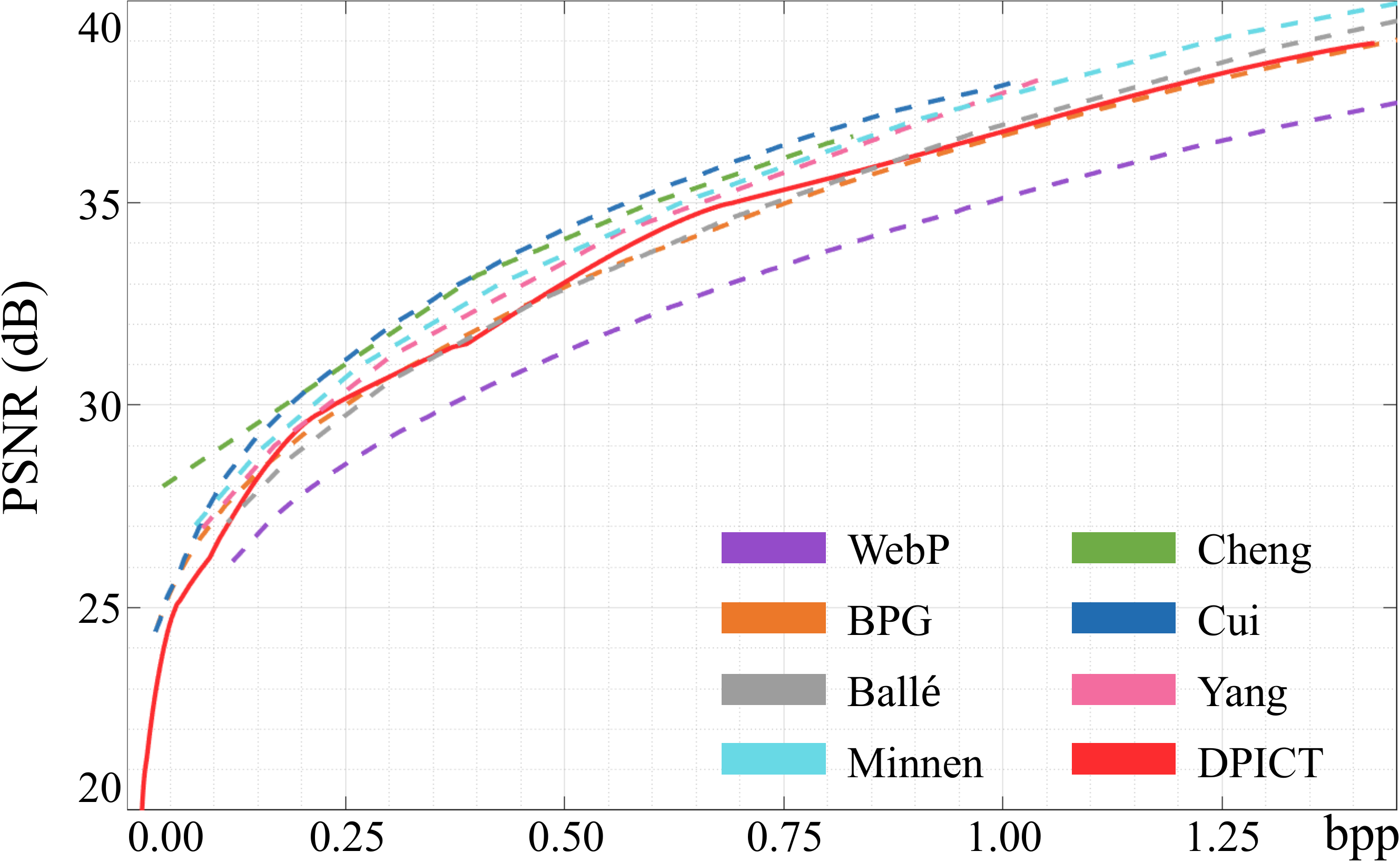}
    \end{center}
    \vspace*{-0.4cm}
    \caption
    {
        RD performance comparison of the proposed DPICT with \textit{non-progressive} codecs on the Kodak dataset: WebP~\cite{WebP}, BPG~\cite{bpg}, Ball{\'e} \etal~\cite{y2018_ICLR_balle}, Minnen \etal~\cite{y2018_NIPS_minnen}, Cheng \etal~\cite{y2020_CVPR_cheng}, Cui \etal~\cite{y2021_CVPR_cui}, and Yang \etal~\cite{y2021_CVPR_yang}.
    }
    \label{fig:comparison_nonprogressive}
\end{figure}

\section{Experimental Results}
\label{sec:experiments}

\subsection{Implementation}

We develop a compression network similar to the Cheng \etal's network \cite{y2020_CVPR_cheng}, but we make some modifications to encode scalable bitstreams. First, we remove the autoregressive mask convolution, which predicts entropy parameters from a latent tensor. The autoregressive prediction assumes that the decoder can reconstruct the same latent tensor $\hat{\mathbf{Y}}$ in the raster scan order as the encoder does. However, in DPICT, the information in $\hat{\mathbf{Y}}$ is progressively transmitted. Therefore, before the decoder reconstructs all trit-planes, the assumption fails and the entropy decoding breaks down. Second, we assume that each latent element $y$ has a Gaussian distribution, instead of a Gaussian mixture model. This is to simplify the MMSE estimation and RD optimization in \eqref{eq:cond_prob} and \eqref{eq:reconstruction_level}$\sim$\eqref{eq:priority}.

A postprocessing network $g_p$ also has the encoder-dec\-oder architecture. It is implemented with residual blocks, attention modules, and subpixel convolution layers \cite{y2020_CVPR_cheng}. However, $g_p$ has 35 layers only, while the encoder $g_a$ and the decoder $g_s$ in the compression network in Figure~\ref{fig:model_schematic_1} have 68 layers. We train two postprocessing networks, targeting at different bit-rates. Note that $\hat{\mathbf{X}}^{n}$ denotes a reconstructed image using the first $n$ trit-planes. When $n$ is not an integer, $\hat{\mathbf{X}}^{n}$ means that, in addition to the first $\lfloor n \rfloor$ trit-planes, $100\times (n-\lfloor n \rfloor)$ \% of the trits in the $(\lfloor n \rfloor+1)$th trit-plane are used for the reconstruction. The first $g_p$ targets at $\hat{\mathbf{X}}^{n}$ for $n \in [0, L-2.9]$, and the second $g_p$ for $n \in (L-2.9, L-1.8]$. We decided these ranges empirically and also observed that the postprocessing is not effective for $\hat{\mathbf{X}}^{n}$, $n \in (L-1.8, L]$, which is already of high quality.

For training, we use the Vimeo90k dataset \cite{y2019_IJCV_xue}. Since it contains frames with overlapping contents, we sample 80,000 frames and randomly crop $256 \times 256$ patches from each sampled frame. We use the Adam optimizer \cite{y2015_ICLR_kingma} with a batch size of 16, a learning rate of $2 \cdot 10^{-5}$, and $\lambda=5$. We perform the scheduled learning according to cosine annealing cycles \cite{y2017_ICLR_huang}. We train the compression network for 200 epochs and then the postprocessing networks for 20 epochs.

For evaluation, we use the Kodak lossless image dataset \cite{kodim} and the CLIC professional validation dataset \cite{CLIC}. The Kodak dataset consists of 24 images of resolution $512 \times 768$ or $768 \times 512$, while the CLIC dataset contains 41 images of higher quality up to 2K resolution. For each image, we determine the number $L$ of trit-planes to cover the values of all elements, after truncating outliers, in $\hat{\mathbf{Y}}_c$. We measure the rate by bits per pixel (bpp), and the distortion by peak-to-signal ratio (PSNR) and multi-scale structural similarity (MS-SSIM)~\cite{y2003_ACS_wang_MS_SSIM}. For MS-SSIM, we convert it to dB scale by $\textrm{MS-SSIM (dB)} = - 10 \cdot \log_{10} (1 - \textrm{MS-SSIM})$.

We conduct all experiments using Pytorch \cite{y2019_NIPS_paszke_pytorch} and CompressAI \cite{y2020_arXiv_begaint_compressai}. Network architecture and implementation details are available in the supplemental document.

\begin{figure}[t]
    \begin{center}
    \includegraphics[width=\linewidth]{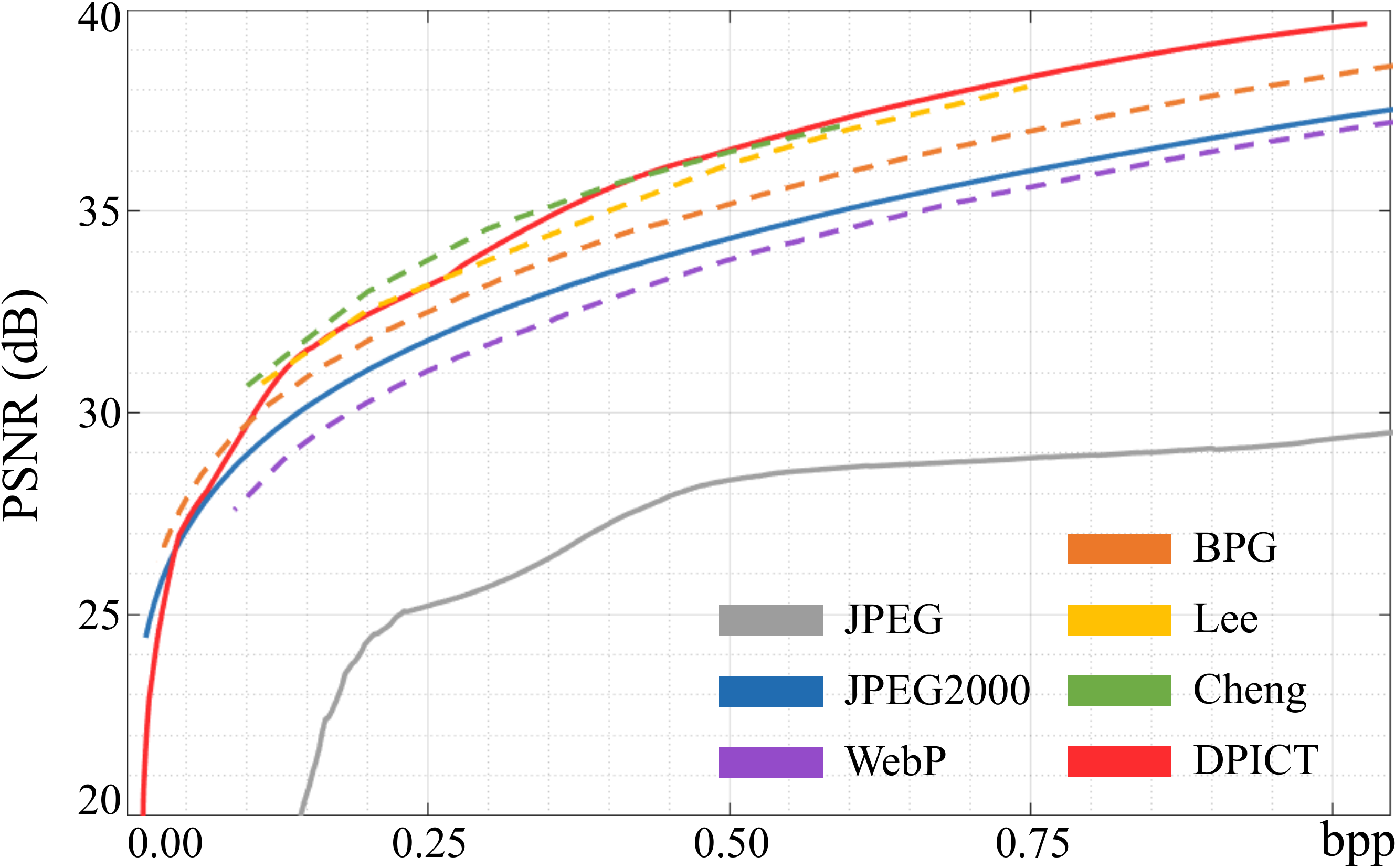}
    \end{center}
    \vspace*{-0.4cm}
    \caption
    {
        RD performance comparison on the CLIC dataset: DPICT is compared with JPEG~\cite{y1992_CE_JPEG}, JPEG2000~\cite{y2001_SPM_JPEG2000}, WebP~\cite{WebP}, BPG~\cite{bpg}, Lee~\etal~\cite{y2019_ICLR_lee}, and Cheng~\etal~\cite{y2020_CVPR_cheng}. Note that only JPEG, JPEG2000, and DPICT are progressive codecs, and their RD curves are solid lines.
    }
    \label{fig:graph_CLIC}
\end{figure}

\begin{figure*}[t]
    \begin{center}
    \includegraphics[width=\linewidth]{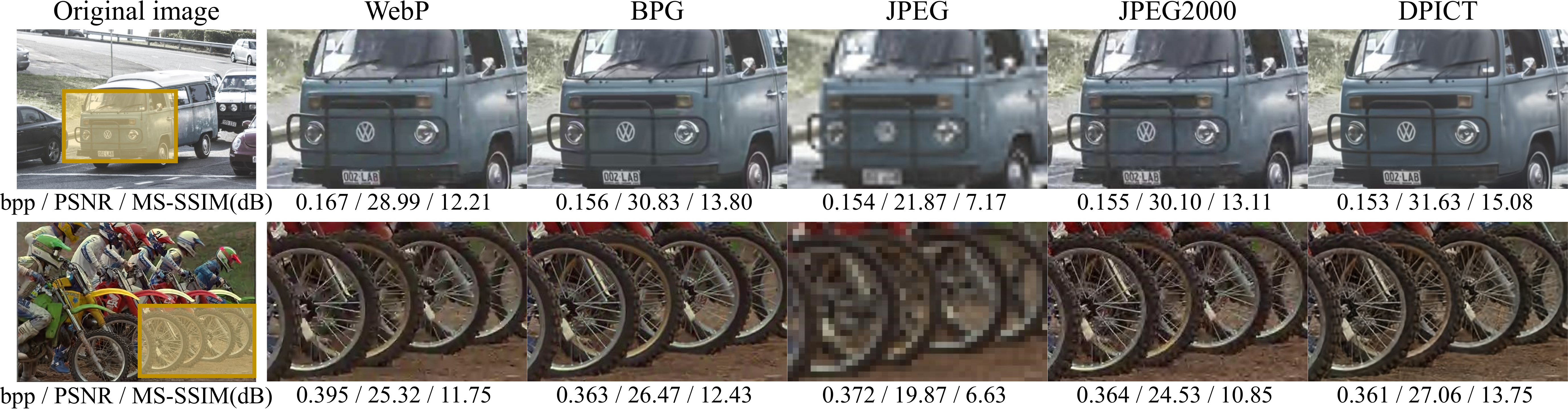}
    \end{center}
    \vspace*{-0.45cm}
    \caption
    {
        Qualitative comparison of reconstructed images at similar rates: WebP \cite{WebP}, BPG \cite{bpg}, JPEG \cite{y1992_CE_JPEG}, JPEG2000 \cite{y2001_SPM_JPEG2000}, and DPICT.
    }
    \label{fig:qualitative_main}
    \vspace*{0.1cm}
\end{figure*}

\begin{figure*}[t]
    \begin{center}
    \includegraphics[width=\linewidth]{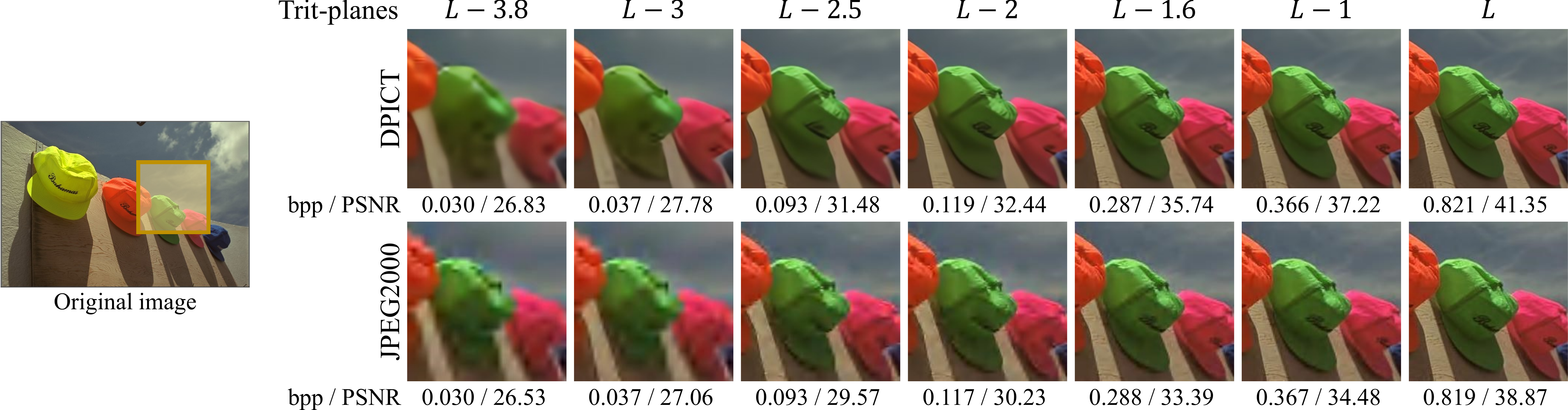}
    \end{center}
    \vspace*{-0.45cm}
    \caption
    {
        Qualitative comparison of progressively reconstructed images at various rates: JPEG2000~\cite{y2001_SPM_JPEG2000} and the proposed DPICT. At the top, the numbers of trit-planes used for the DPICT reconstruction are specified.
    }
    \label{fig:qualitative_progressive}
\end{figure*}

\subsection{Performance comparison}
First, we compare the proposed DPICT with conventional progressive codecs: JPEG~\cite{y1992_CE_JPEG} and JPEG2000~\cite{y2001_SPM_JPEG2000} and learning-based progressive codecs \cite{y2017_CVPR_toderici,y2018_CVPR_johnston,y2021_CVPRW_islam,y2020_DOC_diao,y2020_ICIP_su,y2019_PCS_cai}. Figure~\ref{fig:RD_curve_main} plots the RD curves on the Kodak dataset.
At every rate, DPICT provides the highest PSNR and the highest MS-SSIM, outperforming the second best codecs JPEG2000 and Jhonston \etal~\cite{y2018_CVPR_johnston} considerably. For example, at 0.75bpp, DPICT yields about 1.7dB higher PSNR than JPEG2000, and about 1.1dB higher MS-SSIM than Jonston \etal.

Moreover, it is worth pointing out that DPICT is the only learning-based codec with FGS. The same bitstream of DPICT for an image is reconstructed at 164 different rates, as indicated by red dots in Figure \ref{fig:RD_curve_main}, which can be increased further if needed. On the other hand, Su \etal \cite{y2020_ICIP_su} support coarse granular scalability for four rates only, and their rate range is much narrower than that of DPICT. Cai \etal \cite{y2019_PCS_cai} support two rates only, one for preview images and the other for full-quality images. The RNN-based codecs~\cite{y2018_CVPR_johnston,y2017_CVPR_toderici} support 16 rates, but their PSNR performances are poorer than those of JPEG2000.

Next, Figure \ref{fig:comparison_nonprogressive} compares DPICT with non-progressive codecs \cite{bpg,WebP,y2018_ICLR_balle,y2018_NIPS_minnen,y2020_CVPR_cheng,y2021_CVPR_cui,y2021_CVPR_yang}. Despite the additional functionality of FGS, DPICT outperforms the two traditional codecs, WebP and BPG, and is competitive with the state-of-the-art learning-based codecs.

Figure \ref{fig:graph_CLIC} compares RD curves on the CLIC dataset. We see that DPICT outperforms progressive codecs~\cite{y1992_CE_JPEG,y2001_SPM_JPEG2000} significantly and competes with Cheng \etal \cite{y2020_CVPR_cheng}, which is a state-of-the-art non-progressive codec.

Figure \ref{fig:qualitative_main} compares reconstructed images at similar rates. In areas with complicated texture and sharp edges, such as dirt floor or patterned window, the traditional codecs yield blurring artifacts, but DPICT restores high quality images without noticeable artifacts.

Figure \ref{fig:qualitative_progressive} shows reconstructed images at different rates from a single bitstream. At the top of the figure, we specify how many trit-planes are decoded. At all rates, DPICT provides better qualities than the progressive JPEG2000.

\subsection{Ablation study}

\begin{figure}[t]
    \begin{center}
    \includegraphics[width=\linewidth]{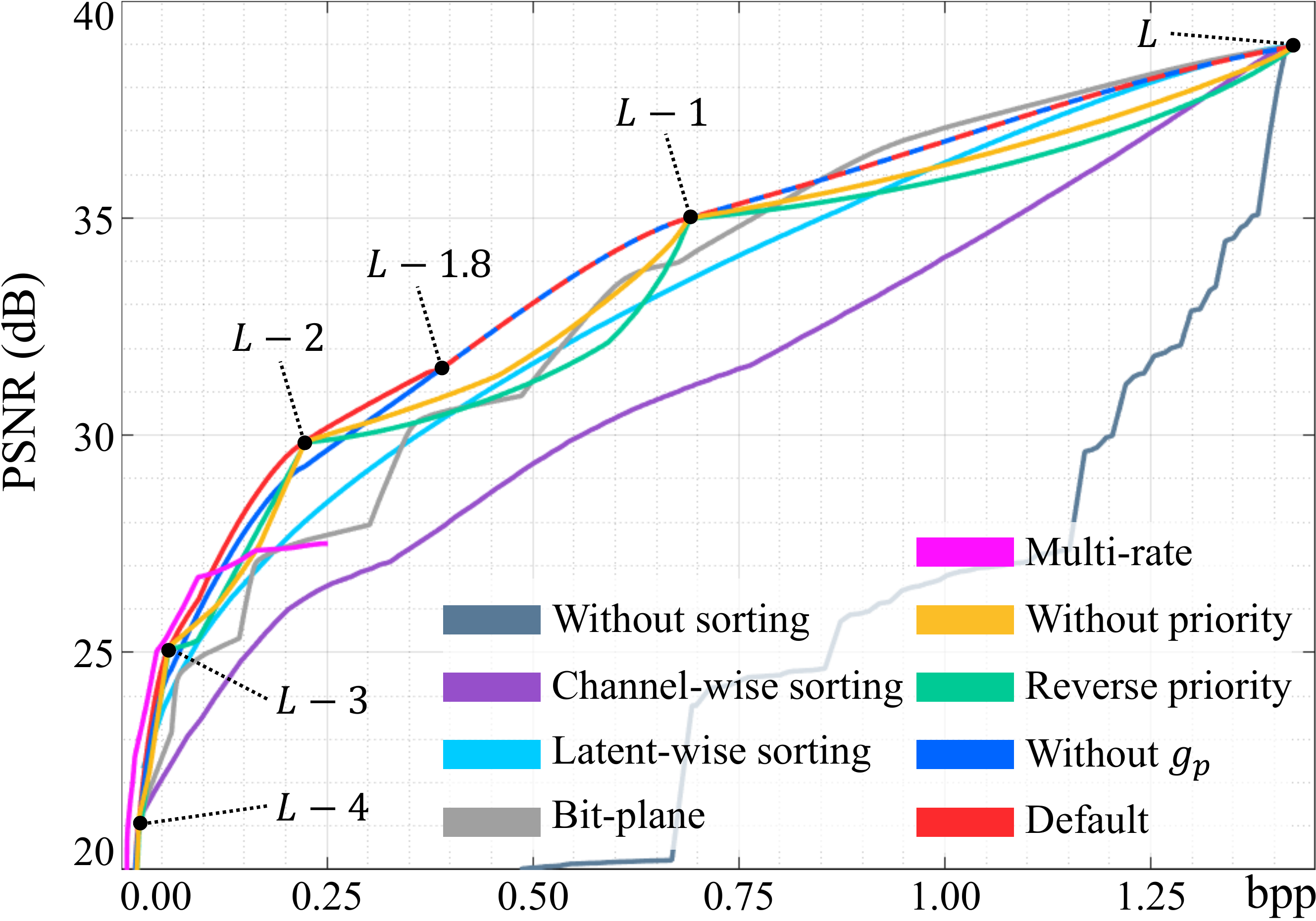}
    \end{center}
    \vspace*{-0.2cm}
    \caption
    {
        Ablation studies of DPICT on the Kodak dataset.
    }
    \label{fig:qualitative_ablation}
\end{figure}

We analyze the compression performance of DPICT in Figure \ref{fig:qualitative_ablation}. Here, each curve represents the result of replacing or removing certain components of DPICT.

First, we compare the proposed DPICT with four baselines excluding the trit-plane coding. In `without sorting,' latent elements are transmitted channel by channel in the raster scan order without priority. This approach underperforms badly. In `channel-wise sorting,' the $C$ channels of the latent tensor  are sorted and transmitted according to the RD priorities.  Instead of channels, `latent-wise sorting' sorts the $K = H \times W \times C$ latent elements. These alternatives can support progressive transmission, but are much inferior to DPICT. `Bit-plane' is the result of replacing the trit-plane coding with the bit-plane coding. As mentioned in Section~\ref{ssec:trit_plane_coding}, $\hat{y}_c^n$ cannot be reconstructed to 0 in the bit-plane coding unless $n=L$. This causes significant performance degradation at low rates, \ie when a bitstream is partially decoded. These results indicate that the trit-plane coding is an essential component of DPICT.

Second, we change the training schema. `Multi-rate' is the result of changing the loss function so that the compression network is trained for multiple rates. Specifically, we replace the loss function in~\eqref{eq:loss_function} with
\begin{equation}
\textstyle
\ell =  \sum_{n=L-3}^{L} \left( \ell_{D} ( \mathbf{X}, \tilde{\mathbf{X}}^n ) + \lambda_n \ell_{R}( \tilde{\mathbf{Y}}^n, \tilde{\mathbf{Z}}; \tilde{\mathbf{M}}, \tilde{\mathbf{\Sigma}} ) \right)
\label{eq:loss_function2}
\end{equation}
to consider partially reconstructed $\tilde{\mathbf{X}}^{L-3}$, $\tilde{\mathbf{X}}^{L-2}$, $\tilde{\mathbf{X}}^{L-1}$, as well as fully reconstructed $\tilde{\mathbf{X}}^{L} = \tilde{\mathbf{X}}$. Note that loss functions for different rates are also combined in \cite{y2021_CVPR_cui,y2021_CVPR_yang}. However, `multi-rate' severely narrows the range of supported rates and is effective only at low rates.

Third, we replace the prioritized transmission. `Without priority' is the result of not using the RD priority in~\eqref{eq:priority}. In this case, trits in each trit-plane are transmitted in the raster scan order. We see that the prioritized transmission is essential for reconstructing high quality $\hat{\mathbf{X}}^n$ when $n$ is not an integer. `Reverse priority' transmits the trits in a trit-plane in the increasing order of the RD priorities. It yields even poorer performance than `without priority.'

\begin{figure}[t]
    \begin{center}
    \includegraphics[width=\linewidth]{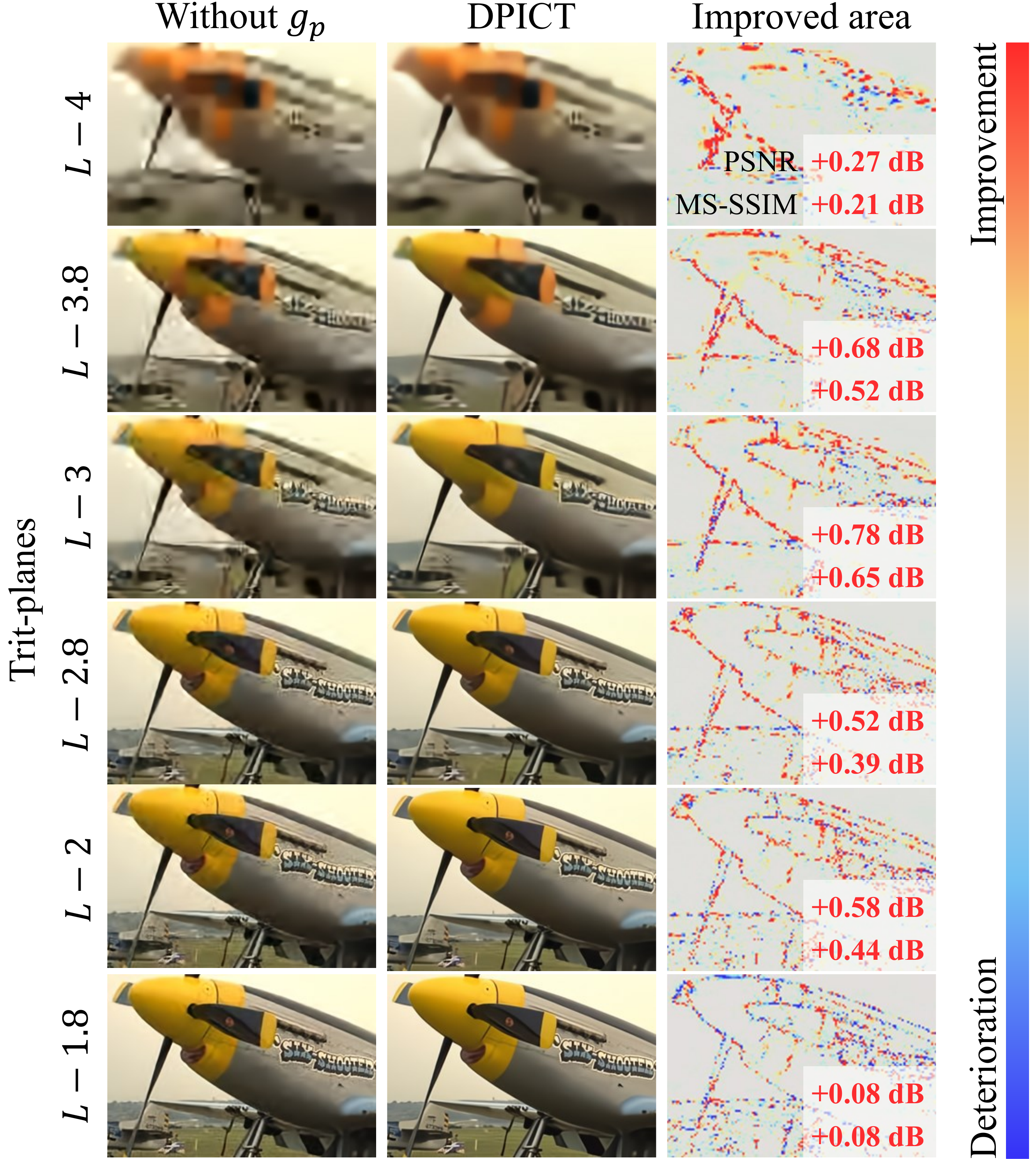}
    \end{center}
    \vspace*{-0.2cm}
    \caption
    {
        Comparison of reconstructed images before and after the postprocessing. Improvement maps are provided, where red and blue means improvement and deterioration, respectively.
    }
    \label{fig:qualitative_addon}
\end{figure}

Last, `without $g_p$' is the result of not using the postprocessing networks. Note that the postprocessing networks improve the quality of a reconstructed image when the rate is lower than about 0.4bpp. Figure \ref{fig:qualitative_addon} shows the impacts of the postprocessing networks  $g_p$. For easier comparison, improvement maps are also provided. When $L-2$ or fewer trit-planes are used, the postprocessing networks consistently improve reconstructed images both qualitatively and quantitatively. On the other hand, for reconstructed images using more than $L-2$ trit-planes, they do not provide clear improvements. Thus, we use the postprocessing networks when fewer than $L-1.8$ trit-planes are decoded.

We provide more experimental results and analysis in the supplemental document. Also, we describe how to implement the proposed DPICT algorithm efficiently in~\cite{y2022_arXiv_jeon}.

\section{Conclusions}
\label{sec:conclusions}
In this paper, we proposed the DPICT algorithm supporting FGS. In DPICT, an image is transformed into a latent tensor using an analysis network. The latent tensor is then represented in a ternary number system and is encoded trit-plane by trit-plane in the decreasing order of significance. Furthermore, in each trit-plane, the trits are transmitted in the decreasing order of the RD priorities. We also developed the post-processing networks to reduce artifacts due to quantization errors when fewer trit-planes are used. Experiments demonstrated that the proposed DPICT provides state-of-the-art performance by outperforming other progressive codecs signficantly.

\section*{Acknowledgments}
This work was supported by the National Research Foundation of Korea (NRF) grants funded by the Korea government (MSIT) (No.~NRF-2021R1A4A1031864 and No.~NRF-2022R1A2B5B03002310).

\newpage

{\small
\bibliographystyle{ieee_fullname}
\bibliography{04515}
}

\end{document}